\documentclass[[aps,prl,reprint,superscriptaddress,floatfix]{revtex4-1}
\usepackage{graphicx}

\usepackage{amsmath}
\usepackage{amssymb}
\usepackage{hyperref}
\usepackage{bm}

\usepackage{color}

\begin{document}

\title{Anisotropic 2D materials for tunable hyperbolic plasmonics}%

\author{Andrei Nemilentsau}
\affiliation{Department of Electrical Engineering \& Computer Science, University of Wisconsin-Milwaukee, Milwaukee, WI 53211, USA}
\author{Tony Low}
\email{tlow@umn.edu}
\affiliation{Department of Electrical \& Computer Engineering, University of Minnesota, Minneapolis, MN 55455, USA}

\author{George Hanson}
\affiliation{Department of Electrical Engineering \& Computer Science, University of Wisconsin-Milwaukee, Milwaukee, WI 53211, USA}

\date{\today}%

\begin{abstract}
Motivated by the recent emergence of a new class of anisotropic 2D materials, we examine their electromagnetic modes and demonstrate that a broad class of the materials can host highly directional hyperbolic plasmons. Their propagation direction can be manipulated on-the-spot by gate doping, enabling  hyperbolic beams reflection, refraction and bending. The realization of these natural 2D hyperbolic media opens up a new avenue in dynamic control of hyperbolic plasmons not possible in the 3D version.
\end{abstract}

\maketitle

A hyperbolic material (HM) is a highly anisotropic material, with real parts of principal components of its permittivity tensor having opposite signs\cite{Poddubny2013,Ferrari2015}. The name of the material is derived from the topology of the $k$-surface which is a hyperboloid (unlike an ellipsoid in conventional anisotropic materials\cite{Kong1986}). Due to this unique topology, waves with very large (theoretically infinite) wave vectors can propagate inside HMs. Existence of the high-$k$ modes means very large photonic density of states inside the material and thus these materials provide unique capabilities for controlling properties of light emitters\cite{Cortes2012}, creating broadband thermal sources\cite{Guo2013}, supporting directional surface plasmons\cite{Alu2015,Yermakov2015}, etc. All HMs known so far, both natural\cite{Basov2014,Novoselov2014,Zhou2014,Novoselov2015,Low2015} and artificial \cite{Poddubny2013,Ferrari2015}, are non-tunable materials with optical response predefined by the design. The hyperbolic regime in metamaterials and metasurfaces relies entirely on geometrical arrangement of the material constituents, and thus is extrinsic in nature. As the geometry of a metamaterial is difficult to modify after the material is created, its optical response at a given frequency can not be changed on-the-spot.  

In this letter we argue that the emerging class of anisotropic 2D materials \cite{zhang2015} is naturally suited to serve as HMs capable of supporting tunable hyperbolic plasmons. The family of 2D materials is very broad (encompassing many elements from the periodic table), with tremendous variety in electronic and optical properties, including metals, semimetals, semiconductors and dielectrics\cite{Miro2014}. Recently, an emerging class of 2D materials with anisotropic electronic and optical properties, such as group V mono- and multi- layers, most notably black phosphorus (BP)\cite{Ye2014,Yuanbo2014,Neto2014,Qiao2014,Low2014,Low2014a,JingboLi2014,Chongwu2015}, the 1T phase of transition metal dichalcogenides (TMDs) \cite{Tongay2014,Tongay2015,Heintz2015} and trichalcogenides\cite{Gomez2014,Cheng2015,Peeters2015a} are garnering attention. Even though these materials are semiconductors with a bandgap varying from the mid-IR to the optical range, injection of free carriers is possible by doping. Doping can be controlled by a gate, allowing for maximum carrier concentrations up to $10^{14}$ cm$^{-2}$\cite{Kim2010}. 

We offer a novel physical mechanism, inherent in nature, for creating highly-tunable two-dimensional hyperbolic materials based on the anisotropic 2D materials in the far- and mid- infrared frequency range. This mechanism relies on the interplay between two types of electron motions defining the electromagnetic response of typical semiconductors; motion of electrons (holes) inside the conduction (valence) band of the material (intraband motion), and electron transitions from valence to conduction band of the material (interband transitions). In isotropic material predominance of intraband motions (interband transitions) renders capacitive (inductive) response. We argue, that in novel highly anisotropic 2D materials, anisotropic electron motion results in inductive material response along one of the optical axis, and capacitive response along the other axis, thus effectively rendering a hyperbolic 2D material. As the intraband motion of electrons in 2D materials can be tuned on-the-spot by applying an electric bias, so does the optical response of the resulting 2D hyperbolic material. Moreover, due to its inherent nature, the hyperbolic regime can be implemented using a single sheet of 2D material without requiring complicated pattering of the substrate. We discuss properties of these hyperbolic plasmons and examine their existence across the material parameter space. 

We begin by considering a minimal model for the conductivity tensor for an anisotropic semiconducting media, which accounts for both the intraband electron motions
(first term on the right hand) and for interband electron transitions (second term), 
\begin{equation} \label{Eq:cond}
\underline{\underline{\sigma}} = 
\left(
	\begin{array}{cc}
		\sigma_{xx} & 0 \\
		0 & \sigma_{yy}
	\end{array}
\right),
\end{equation} 
where 
\begin{equation} \label{Eq:cond_comp}
\sigma_{jj} = \frac{i e^2 }{\omega + i\eta} \, \frac{n}{m_j} + s_j \left[ \Theta(\omega - \omega_j) + \frac{i}{\pi}  \ln\left|\frac{\omega - \omega_j}{\omega + \omega_j} \right|\right],
\end{equation}
$j=x,y$, where $n$ is the concentration of electrons, $m_{j}$  is the electron's effective mass along the $j$th direction, $\eta$ is a relaxation time, $\omega_j$ is the frequency of onset of interband transitions for the $j$th component of conductivity, and $s_j$ accounts for the strength of the interband component.  It may seem somewhat counter intuitive that the $x$ and $y$ components of the conductivity tensor can have different frequencies of inter-band transitions. However, this is the case in BP multi-layers, where different crystallographic directions have different symmetries, and thus the dipole moment for electron interband transitions is zero along one of the directions\cite{Qiao2014}. 

We use the Drude model in order to calculate intraband conductivity. Interband conductivity is introduced phenomenologically by defining absorption through a step function $\Theta(\omega-\omega_j)$. The imaginary part of interband conductivity then follows from the Kramers-Kronig relations. In the general case, parameters in Eq. \ref{Eq:cond_comp} should be either measured experimentally or estimated theoretically using, for example, the Kubo formula. Assuming that material parameters are isotropic, Eq. \ref{Eq:cond_comp} can describe the conductivity of graphene\cite{Koppens2011}. On the other hand, in the anisotropic case this expression can be used as an approximation for the conductivity of multi-layer black phosphorus\cite{Low2014,Low2014a}.

\begin{figure}
	\centering
	\includegraphics[width=\linewidth]{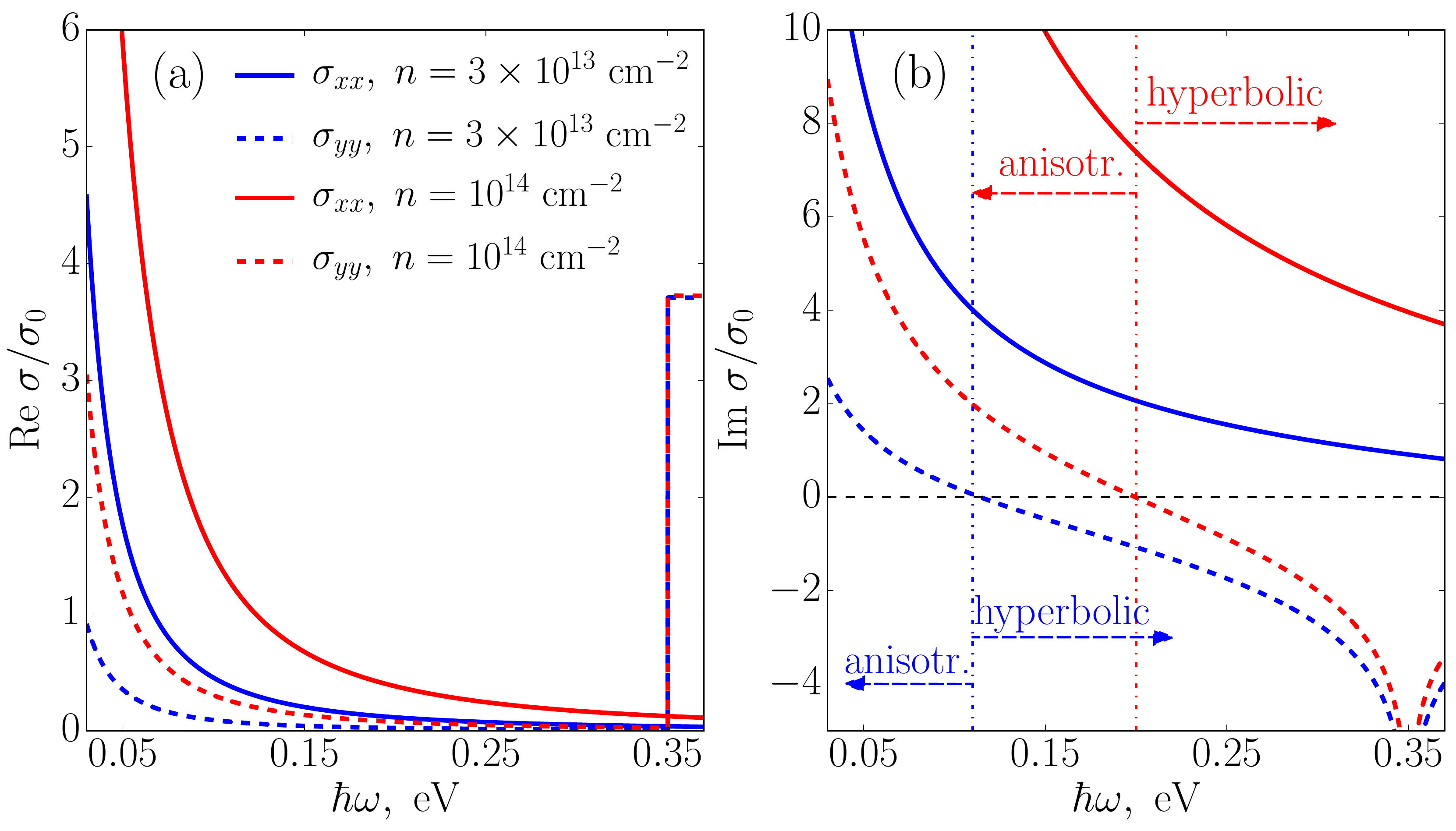}
	\caption{Real (a) and imaginary (b) parts of conductivity of 2D material. Parameters of 2D material are $m_x = 0.2 m_0$, $m_y = m_0$,  $\eta = 0.01$ eV, $w_x = 1$ eV, $s_x = 1.7s_0$, $w_y = 0.35$ eV, and $s_y = 3.7s_0$. $s_0 = e^2/4\hbar$, $m_0$ is the free-electron mass. }
	\label{fig0} 
\end{figure}

We start our consideration with a particular set of parameters (see Fig. \ref{fig0}) and then study how the material response changes when one or several parameters vary.  The conductivity of a 2D material is presented in Fig. \ref{fig0}. In general, two distinct regimes in the electromagnetic response of 2D material can be identified. If frequency is sufficiently low (to the left of dash-dotted red or blue lines in Fig. \ref{fig0}a), then conductivity is of a pure Drude type ($\mathrm{Im}\sigma > 0$) and we refer to the material as purely anisotropic (in order to distinguish from hyperbolic anisotropic regime). If frequency is sufficiently high the contribution from interband electron transitions may become dominant, and the imaginary part of conductivity becomes negative. Due to the crystal asymmetry, $\mathrm{Im}\sigma_{xx}$ and $\mathrm{Im} \sigma_{yy}$ can in principle change sign at different frequencies and a frequency interval where $\mathrm{Im}\sigma_{xx} \cdot \mathrm{Im} \sigma_{yy} < 0$ exists, which indicates the hyperbolic regime (to the right of dash-dotted red or blue lines). Such natural hyperbolicity in anisotropic 2D materials is unlike hyperbolic metamaterials that have been extensively studied so far\cite{Poddubny2013,Alu2015}. Moreover, 2D materials offer unique opportunity of switching between the two different plasmonic regimes on-the-spot by applying an electrical bias (compare two doping cases in Fig. \ref{fig0}b). 

\begin{figure}
\centering
\includegraphics[width=\linewidth]{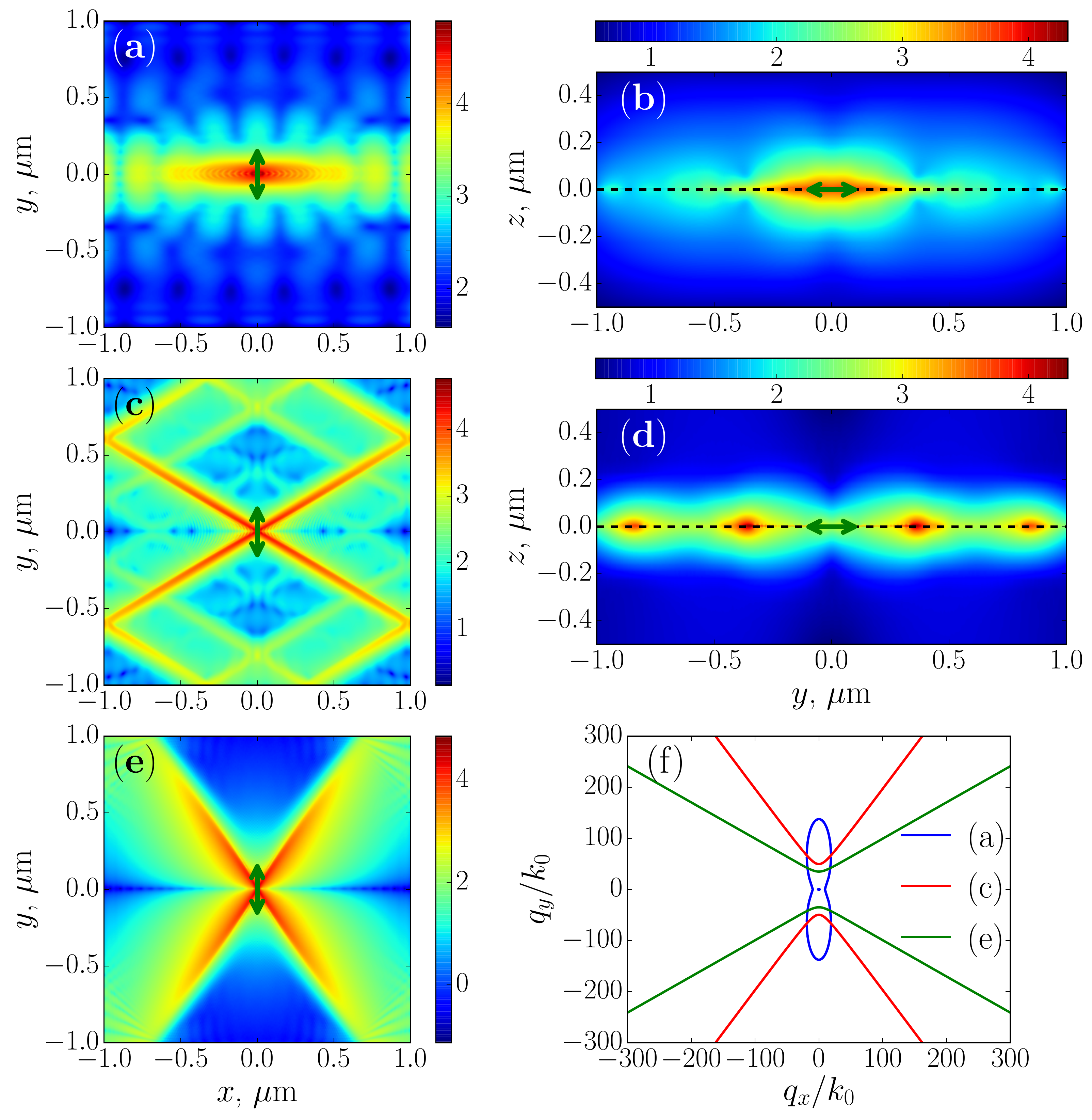}
\caption{(a)-(e) Spatial distribution (log scale) of electric field, $|\mathbf{E}|$, of a surface plasmon excited by $y$-polarized electric dipole (green arrow in panels (a)-(e)) in a $2\times2$ $\mu$m sheet of 2D material placed in plane $z = 0$ (dashed black line in panels (b,d)). Distribution of electric field is calculated in planes $z = 10$ nm (a,c,e)  and $x = 0.6$ $\mu$m (b,d). (a,b) $n = 10^{14}$ cm$^{-2}$, $\hbar\omega = 0.165$ eV, (c,d) $n = 10^{14}$ cm$^{-2}$, and $\hbar\omega = 0.3$ eV, (e) $n = 3\times10^{13}$ cm$^{-2}$, $\hbar\omega = 0.3$ eV. (f) k-surfaces, $\omega(q_x,q_y)=\mathrm{const}$, for plasmons in panels (a)-(e). }
\label{fig1} 
\end{figure}

We study properties of plasmons in pure anisotropic and hyperbolic regimes by exciting surface waves in a $2\mu$m$\times2\mu$m sheet of 2D material by a $y$-polarized electric dipole placed on top of the material (see Fig. \ref{fig1} for details). We choose material parameters to be the same as in Fig. \ref{fig0}. Hereinafter we assume that the medium above and below the 2D material is vacuum. The distribution of plasmon electric field (obtained solving Maxwell’s equations numerically using a commercial finite-difference time-domain method (FDTD) from Lumerical\cite{Lumerical}) is presented in Figs. \ref{fig1}a-e. Figs. \ref{fig1}a,b demonstrate the plasmon field intensity, $\mathbf{E}$, in the purely anisotropic regime. In this case the plasmon propagates along one of the crystallographic axes of the material ($x$-axis in this case). Figs. \ref{fig1}c-e show plasmon propagation in the hyperbolic regime. Plasmon energy is channeled as very narrow beams, the angle between direction of beams propagation and $x$-axis being a function of electron concentration $n$. Thus, the direction of plasmon propagation in the hyperbolic regime can be controlled on-the-spot by biasing the 2D material. This opens a new avenue for solving the problem of plasmon guiding, which in conventional materials requires structuring of the material geometry \cite{Moreno2008} or applying external magnetic fields\cite{Steinberg2010,Steinberg2012}.

In order to understand the behavior of surface waves it is instructive to inspect the plasmon dispersion relation in 2D materials. Considering only eigenmodes bounded to the 2D material plane,  $e^{i(q_x x + q_y y) } e^{\pm p z}$ (for $z \lessgtr 0$), we obtain\cite{Hanson2008}
\begin{align}
\left(q_x^2 - k_0^2\right) \sigma_{xx} & + \left(q_y^2 - k_0^2\right) \sigma_{yy} =   2 i p \omega \left(\varepsilon_0 + \frac{\mu_0 \sigma_{xx} \sigma_{yy}}{4} \right) , \label{Eq:dispersion}
\end{align}
where $p = \sqrt{q_x^2 + q_y^2 - k_0^2}$, $k_0 = \omega \sqrt{\varepsilon_0 \mu_0}$ is the vacuum wavenumber, and $\varepsilon_0$, $\mu_0$ are  vacuum permittivity and permeability, respectively. Solution of Eq. \eqref{Eq:dispersion}, $\mathbf{q}(\omega)$, defines plasmon wavefront propagation. However, in anisotropic materials in the general case directions of wavefront and energy propagation do not coincide. Instead, it is group velocity $\mathbf{v}_{g} = \nabla_{\mathbf{q}}\omega(\mathbf{q})$ that is in the direction of plasmon energy propagation\cite{Kong1986}. Analytical computation of group velocity requires the explicit form of $\omega(\mathbf{q})$, which is not available in our case. Instead, we use $k$-surfaces, $\omega(q_x,q_y) = \mathrm{const}$, for further analysis (see Fig. \ref{fig1}f). Group velocity, and thus direction of the plasmon energy flow, has to be orthogonal to the $k$-surface. Assuming that conductivity is purely imaginary $\sigma_{jj} = i\sigma''_{jj}$ (losses are small in our system, see Fig. \ref{fig0}), and that $q_x, q_y \gg k_0$, Eq. \eqref{Eq:dispersion} can be further simplified
\begin{align}
\frac{q_x^2}{\sigma''_{yy}}  + \frac{q_y^2}{\sigma''_{xx}}  = 2 p \omega \left(\frac{\varepsilon_0}{\sigma''_{xx} \sigma''_{yy}} - \frac{\mu_0 }{4} \right). \label{Eq:dispersion1}
\end{align}

In the purely anisotropic case ($\sigma''_{xx}, \sigma''_{yy} > 0$), if the right side were constant, Eq. \eqref{Eq:dispersion1} would be an equation for an ellipse in $\mathbf{q}$-space with the axis oriented along $q_x$ and $q_y$. The length of the ellipse's principal axes along $q_x$($q_y$) direction is proportional to $\sigma''_{yy}$($\sigma''_{xx}$). Thus, the $k$-surface always takes quasi-eliptic form elongated along the direction of the smallest component of the conductivity tensor, the degree of elongation being defined by the ratio of $\sigma''_{xx}$ and $\sigma''_{yy}$.  This is what we see in Fig. \ref{fig1}f (blue line), where $\sigma''_{xx} = 0.55$ mS, $\sigma''_{yy} = 0.039$ mS. Due to the strong elongation of the $k$-surface along the $q_y$-axis, group velocity for predominant range of plasmon momentum $\mathbf{q}$ points approximately along the $q_x$ axis, which causes plasmons to carry energy in the direction of the $x$ crystallographic axis. If the ratio of $\sigma''_{xx}$ and $\sigma''_{yy}$ decreases, the plasmon beam spreads. Finally, in the case of $\sigma''_{xx} = \sigma''_{yy}$ the $k$-surface turns into circle and there is no preferential direction for energy transfer.  

In the hyperbolic case ($\sigma''_{xx}\cdot\sigma''_{yy} < 0$) the $k$-surface is a hyperbola (see Fig. \ref{fig1}, red and green lines), with the hyperbola asymptotes being defined by $q_y = \pm q_x \sqrt{|\sigma''_{xx}/\sigma''_{yy}|}$ (see Eq. \eqref{Eq:dispersion1}). Normals to the asymptotes always point along the same direction defined by 
\begin{equation} \label{Eq:direction}
y = \pm x \sqrt{|\sigma''_{yy}/\sigma''_{xx}|}
\end{equation}
which is also the direction of the plasmon beams. As plasmons with all momenta carry energy along the same direction, the beams are significantly more localized than in the purely anisotropic case. Moreover, as can be seen from Eq. \eqref{Eq:direction}, beams are leaning towards the crystallographic axis with the largest conductivity. From Fig. \ref{fig0} it is clear that the conductivity of the 2D material, and thus the direction of plasmon beam propagation, can be changed on-the-spot by doping. Indeed, in the case of Fig. \ref{fig1}c ($n=10^{14}$ cm$^{-2}$) $\sigma''_{xx} = 0.3$ mS, $\sigma''_{yy} = -0.11$ mS, which corresponds to the propagation angle of 31$^{\circ}$, while in the case presented in Fig. \ref{fig1}e ($n=3\times10^{13}$ cm$^{-2}$) $\sigma''_{xx} = 0.076$ mS, $\sigma''_{yy} = -0.15$ mS, and propagation angle is 55$^{\circ}$. It should be noted that the propagation angles obtained using Eq. \eqref{Eq:direction} agree with the results of numerical simulations presented in Fig. \ref{fig1}. Finally, we want to point out that the beam can experience multiple reflections from the edges of a 2D material sheet, as can be clearly seen from Fig. \ref{fig1}c. This leads us to the discussion of \textit{ray optics of the hyperbolic surface plasmons}, reminiscent of conventional geometrical optics, in the later part of the paper. 

It should be noted that there is nothing special about the choice of parameters in Figs. \ref{fig0},\ref{fig1}. Anisotropic and hyperbolic plasmons are supported by 2D materials with a broad range of parameters. It is quite obvious that as long as the material is anisotropic and has free electrons, pure anisotropic plasmons can propagate in this material when the frequency is sufficiently low. In fact, multi-layer BP is an example of such a 2D material that is capable of guiding anisotropic plasmons\cite{Low2014}. 

\begin{figure}
	\centering
	\includegraphics[width=\linewidth]{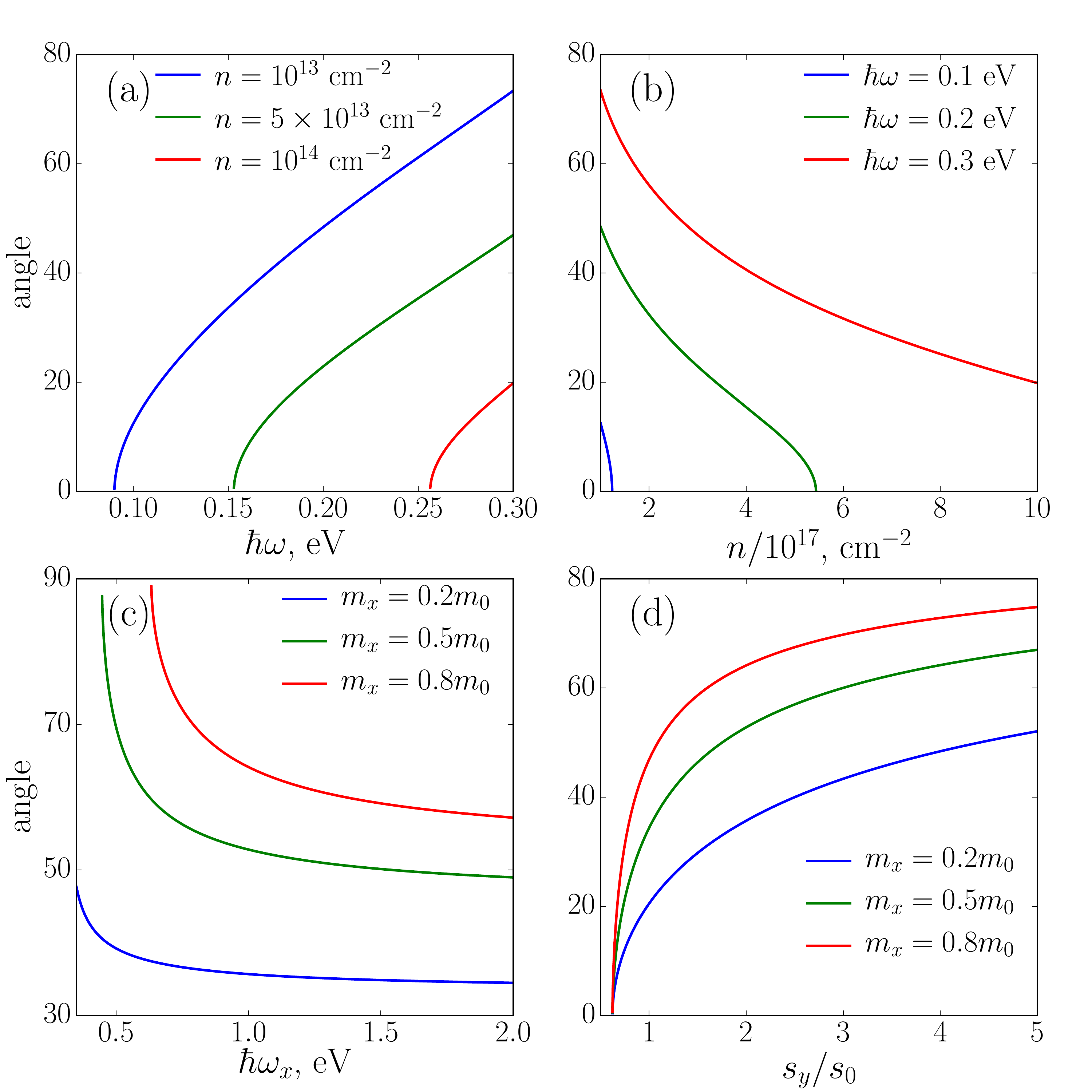}
	\caption{Angle between plasmon beam direction and $x$-axis, for $\eta = 0.01$ eV, $m_x = 0.2 m_0$, $m_y = m_0$, $s_x = s_y = 2s_0$,  $\hbar\omega_x = 1$ eV, $\hbar\omega_y = 0.35$ eV, $n = 5\times10^{13}$ cm$^{-2}$, and $\hbar\omega$ = 0.3 eV, unless stated otherwise in the figure.}
	\label{fig2} 
\end{figure}

Occurrence of a hyperbolic regime requires an interplay between electron intraband motion and interband transitions, as it is interband transitions that cause the imaginary part of conductivity to change sign. Thus, the hyperbolic regime comes into play at frequencies sufficiently close to the frequency of interband transitions. In Fig. \ref{fig2} we clarify this notion of closeness by considering properties of hyperbolic plasmons for a broad range of parameters. We assume that $m_x < m_y$, which means that the $x$-component of conductivity is predominantly Drude-like, while the $y$-component of conductivity is the one that is mostly affected by interband motions of electrons. 

In Figs. \ref{fig2}a,b we inspect the behavior of surface plasmons at different frequencies and electron concentrations. One can see that the direction of plasmon beams leans towards the $x$-axis at small frequencies and/or high concentrations. This is due to large values of $\sigma''_{xx}$, which is predominantly Drude-like (see Eq. \eqref{Eq:direction}), compared to $\sigma''_{yy}$. Moreover, when the frequency is very low and electron concentration is very high, intraband motion of electrons dominates $\sigma''_{yy}$ as well. This makes the material purely anisotropic (thus the cutoff in Figs. \ref{fig2}a,b). On the other hand, at low concentrations and high frequencies, interband motion of electrons is dominant. Conductivity $|\sigma''_{yy}|$ exceeds $|\sigma''_{xx}|$ and  plasmon beams propagate along directions close to the $y$-axis. 

Figs. \ref{fig2}c,d demonstrate the behavior of plasmon beams at a given frequency and electron concentration, for  different values of $m_x$, $\omega_x$ and $s_y$. One can see that any change in the set of parameters (decrease of $m_x$, increase of $\omega_x$) that leads to the dominance of intraband motions of electrons over the interband ones shifts directions of plasmon beams towards the $x$ axis. When $\omega_x$ becomes high, the contribution of interband motions to $\sigma''_{xx}$ is negligible, and further increase of $\omega_x$ does not change the propagation angle, thus the saturation shown in Fig. \ref{fig2}c. On the other hand, the contribution of interband transitions to the conductivity increases with an increase in $s_y$, which explains the shift of the plasmon beam propagation direction towards the $y$-axis. In general, we can see that the hyperbolic regime in 2D materials is robust under a very broad range of experimentally achievable parameters. 

Due to their outstanding mechanical properties, 2D materials can withstand a significant amount of strain, which provides unique capabilities to vary material parameters in Eq. \eqref{Eq:cond_comp} in a wide range using strain engineering\cite{Guinea2015}. Indeed, significant variations in the band gap (between 0.38 eV and 2.07 eV) and electron effective mass (up to $4m_0$, and relative magnitudes of $m_x$ and $m_y$ can be interchanged) in BP under the influence of strain\cite{Guo2014,Yang2014,Peeters2015} and/or out-of-plane static electric field\cite{JingboLi2014} were theoretically predicted and experimentally demonstrated\cite{Su2015}. Modification of electronic and optical properties of ReSe$_2$\cite{Tongay2015}, TiS$_3$\cite{Peeters2015a} under the influence of mechanical deformations was also reported. Another potential way of engineering material parameters consists in alloying 2D materials or creating heterostructures by stacking 2D materials vertically\cite{Geim2013,LiuZheng2015}. 

\begin{figure}
	\centering
	\includegraphics[width=\linewidth]{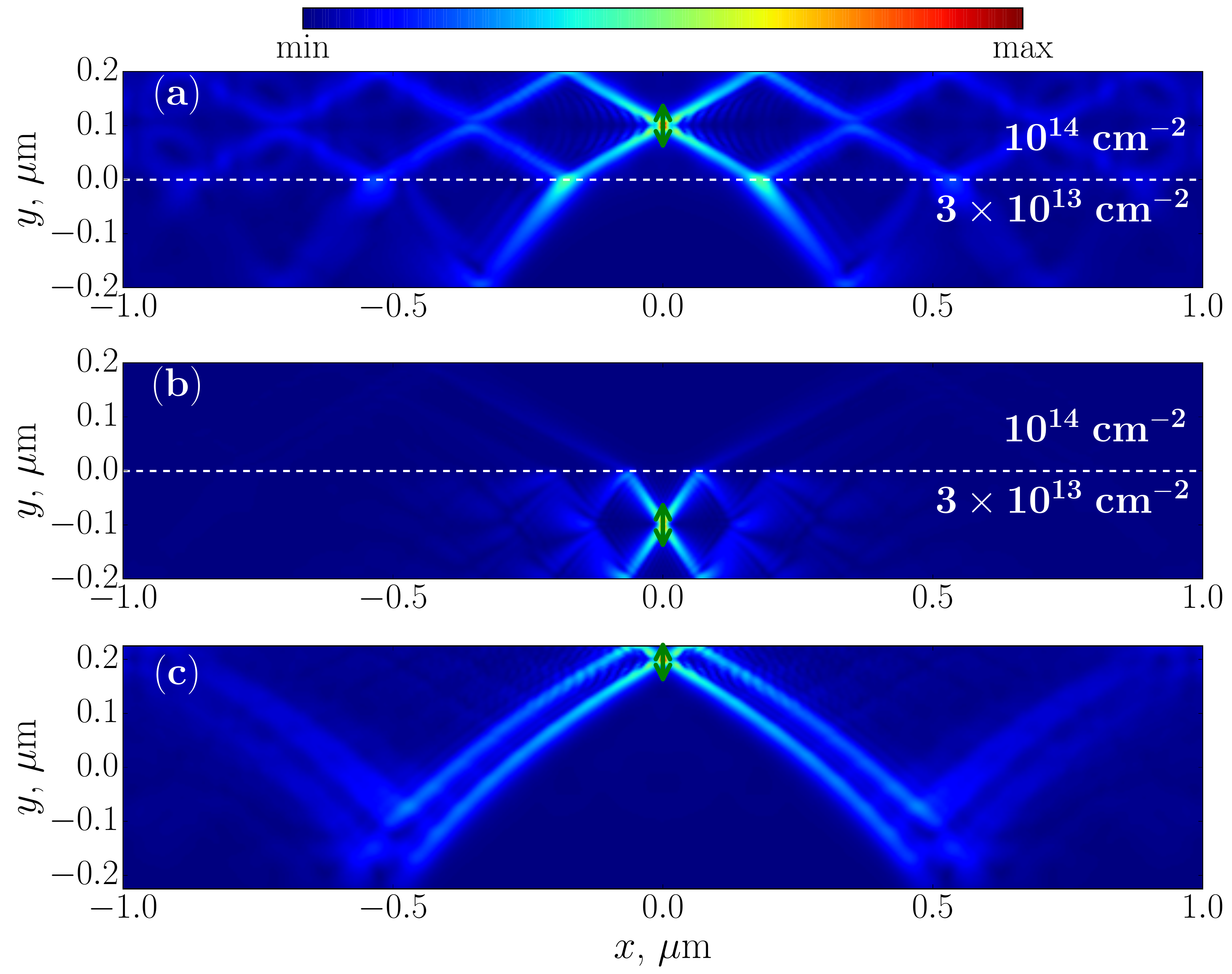}
	\caption{Plasmon launched by a $y$-polarized electric dipole in a layered medium made by placing sheets of 2D materials with different electron concentrations next to each other. Dipole position is designated by a green arrow. Parameters of 2D materials are the same as in Fig. \ref{fig0}. (a,b) Concentration of electrons in the material above dashed white line is $n = 10^{14}$ cm$^{-2}$, while below the white line is $n = 3\times 10^{13}$ cm$^{-2}$. (c) Electron concentration varies gradually from $n = 10^{14}$ cm$^{-2}$ (upper layer) to $n = 2\times 10^{13}$ cm$^{-2}$ (lower layer).}
	\label{fig3} 
\end{figure}

The realization of these natural 2D hyperbolic materials can potentially open up a new avenue in dynamic control of hyperbolic surface waves not possible in the 3D version. In particular, in Fig. \ref{fig3} we consider propagation of plasmon rays through the interface between two sheets of 2D material with different electron concentrations placed next to each other. We can clearly see reflection and refraction of the plasmons at the interface, with reflection angle being equal to the incident angle and refraction angle being defined by the conductivity of the second medium. Moreover, for a plasmon beam incident from the region of higher electron concentration (Fig. \ref{fig3}a) most of the plasmon's energy is transferred through the interface to the lower doped medium. In the opposite case (Fig. \ref{fig3}b), most of the plasmon's energy is reflected back to the lower doped medium. Finally, we implement an analogue of a graded-index plasmonic material by gradually varying the doping across y from $10^{14}$ cm$^{-2}$ to $2\times10^{13}$ cm$^{-2}$(see Fig. \ref{fig3}c). One can see bending of the plasmon ray resembling bending of the light ray in graded-index medium. 

In conclusion, we demonstrated in a general manner for the first time that anisotropic 2D materials can support propagation of highly directional and tunable hyperbolic  surface waves. In the purely anisotropic regime ($\mathrm{Im}\sigma''_{xx} > 0$, $\mathrm{Im}\sigma''_{yy} > 0$) plasmons propagate along the crystallographic axis with larger conductivity, with the ratio of the imaginary parts of conductivities defining the degree of plasmon localization along the axis. In the hyperbolic regime ($\mathrm{Im}\sigma''_{xx}\cdot\mathrm{Im}\sigma''_{yy} < 0$), plasmons propagate as a very localized beam along the direction defined by the ratio of the imaginary parts of the conductivities. Unlike 3D HMs, new 2D natural HMs are highly tunable, the frequency range of the hyperbolic regime as well as the direction of plasmon beams propagation can be modified on-the-spot by gate doping of the 2D material. Through doping control reflection, refraction and bending of plasmon beams can be achieved. By elucidating the existence of an interesting hyperbolic regime due to the interplay of intra- and inter-band electron motion, we hope this will inspire and guide their experimental search and verification.

\bibliography{Directional_plasmon}

\end{document}